\documentclass[final]{svjour3}
\usepackage[dvipdfmx]{graphicx}
\usepackage{rotating}
\usepackage{amssymb}
\usepackage{mathptmx}
\usepackage[numbers, sort&compress]{natbib}
\usepackage{multicol}
\usepackage{subfig}
\usepackage[LGRgreek]{mathastext}
\makeatletter
\journalname{Journal of Low Temperature Physics}

\bibpunct{[}{]}{,}{n}{}{,}

\begin{document}

\newcommand{\hdblarrow}{H\makebox[0.9ex][l]{$\downdownarrows$}-}

\title{AC/DC characterization of a Ti/Au TES with Au/Bi absorber for X-ray detection}

\author{E. Taralli \and. C. Pobes \and  P. Khosropanah \and L. Fabrega \and A. Cam\'on \and L. Gottardi \and K. Nagayoshi \and M. Ridder \and M. Bruijn \and J.R. Gao}

\institute{
E. Taralli \and P. Khosropanah \and L. Gottardi \and K. Nagayoshi \and M. Ridder \and  M. Bruijn \and J.R. Gao\\
SRON Netherlands Institute for Space Research\\Sorbonnelaan 2, Utrecht, 3584 CA, The Netherlands\\
\email{e.taralli@sron.nl}\\
\\
C. Pobes \and A. Cam\'on\\
ICMA, Zaragoza, Spain\\
\email{carlos.pobes@gmail.com}\\
\\
L. Fabrega\\ 
ICMAB-CSIC, Barcelona, Spain\\
\\
J.R. Gao\\
Faculty of Applied Science, Delft University of Technology\\
Lorentzweg 1, 2628 CJ, Delft, The Netherlands\\
}

\maketitle

\begin{abstract}

Transition-edge sensors (TESs) are used as very sensitive thermometers in microcalorimeters aimed at detection of different wavelengths. In particular, for soft X-ray astrophysics, science goals require very high resolution microcalorimeters which can be achieved with TESs coupled to suitable absorbers. For many applications there is also need for a high number of pixels which typically requires multiplexing in the readout stage. Frequency Domain Multiplexing (FDM) is a common scheme and is the baseline proposed for the ATHENA mission. FDM requires biasing the TES in AC at MHz frequencies. Recently there has been reported degradation in performances under AC with respect to DC bias. In order to assess the performances of TESs to be used with FDM, it is thus of great interest to compare the performances of the same device both under AC and DC bias. This requires two different measurement setups with different processes for making the characterization.
We report in this work the preliminary results of a single pixel characterization performed on a TiAu TES under AC and afterwards under DC bias in different facilities. Extraction of dynamical parameters and noise performances are compared in both cases as a first stage for further AC/DC comparison of these devices.\\


\end{abstract}

\section{Introduction}
The Netherlands Institute for Space Research (SRON) is currently developing a Frequency Domain Multiplexing (FDM) readout system as baseline and X-ray TiAu transition-edge sensor (TES) microcalorimeter arrays as a backup technology for the X-ray integral field unit (X-IFU) instrument \cite{xifu} inside the future ATHENA mission \cite{athena} led by ESA and to be launched in 2030s.\\
Our current FDM readout system applies a set of 18 sinusoidal AC carriers (between 1 and 5 MHz), which bias the TES detectors at their working points. This is a small version of the baseline readout for the Flight Model which will consist of channels with 40 pixels.\\
Another technique being developed for the X-ray TES readout as a backup option for X-IFU instrument is Time-Division Multiplexing (TDM) \cite{doriese}, where the important characteristic is that the TESs are DC-biased.\\
In order to fully understand the main differences between these two readout schemes and hence the behaviour of the devices involved in large arrays,  it is worth to probe and compare their functionality and their properties under both DC and AC bias. It is essential to demonstrate that the observed good performance of a single pixel under constant voltage bias are maintained even when the TES works as a modulator \cite{gottardi}.\\
In this paper we present a preliminary comparison by means of IV curves, complex impedance measurements and noise spectra of an X-ray TiAu TES microcalorimeter under DC bias performed at the Institute of Material Science of Aragon (ICMA) and under AC bias at a frequency of 3.5 MHz performed in SRON.\\
 \vspace{-0.8cm}
 \section{Detector and Experimental setups}


\subsection{TES under test}
The sensor pixel for this characterization was fabricated at SRON and was chosen from a 5$\times$5 array with uniform design pixels. It consists of a 140$\times$100 $\mu$m$^2$ TiAu (20/50 nm) bi-layer thermometer with three additional normal metal strips, situated on a 1 $\mu$mm thick SiN membrane. The thermometer has a T$_c $$\sim$100 mK, a normal resistance R$_N$ = 220 m$\Omega$ and a thermal conductance to the bath G$\sim$150 pW/K.  The X-ray absorber, consisting of 3.5 $\mu$m Bi on top of 3 $\mu$m Au, has a size of 248$\times$248 $\mu$m$^2$.
It has, at its corners, four contact points to the membrane and an additional contact point in the centre of the TES. More details on the fabrication of the TES array can be found in Ref. \cite{pourya}.

\subsection{AC bias and setup}
The characterization at SRON has been done in our FDM readout system in single pixel mode, where a TES is biased by a carrier signal with a bias frequency f$_c$ between 1 and 5 MHz. A high-Q superconducting LC resonator filter chip defines the different bias frequencies f$_c$ \cite{Qlc}. This chip contains 18 LC-resonator circuits, with resonance frequencies separated by nominally 200 kHz. Coil inductance L = 400 nH for all 18. The TES current is picked up by a two-stage SQUID assembly, consisting of a low-power single SQUID at the base temperature and a high-power SQUID array at the 2K stage.\\
The TES array chip and the cryogenic components of the FDM readout were mounted in a low magnetic impurity copper bracket fitted into an Al shield and accommodated in a dilution refrigerator with a bath temperature T$_{bath}$$\sim$40 mK. T$_{bath}$ on the bracket can be locally tuned  by a heater and a thermometer directly connected to the setup itself. A Helmholtz coil placed on top of the array is used to apply an uniform perpendicular magnetic field on the TES array.\\
\vspace{-0.7cm}
\subsection{DC bias and setup}
The Council for Scientific Research in Spain (CSIC) is also involved in the development of TES sensors based on Mo/Au proximity bilayers with Au/Bi absorbers (developed by the Institute of Material Science of Barcelona ICMAB and the Institute of Material Science of Aragon - ICMA). 
DC characterization at ICMA is performed in an Oxford dilution refrigerator with a T$_{bath}$$\sim$30 mK. The TES chip is attached to the mixing chamber and the experimental volume is shielded against external magnetic fields by a mu-metal with a lead layer inside. The holder hosts a compensating coil to cancel remnant magnetic fields although it could not be properly used in these measurements. TES current can be measured through a two-stage low noise SQUID (manufactured at the PTB Institute in Berlin) with 2 m$\Omega$ shunt resistor.\\
TES is biased by means of a DC current source from the same Magnicon electronics that controls the SQUID and TES polarization. All the measurements were performed in Flux-Locked-Loop (FLL) mode with a feedback resistance of 100 k$\Omega$. \\
\vspace{-0.8cm}

\section{Fitting of Complex Impedance}

\begin{figure}[htbp]
\begin{center}
	\subfloat[]{\includegraphics[width=0.4\linewidth, keepaspectratio]{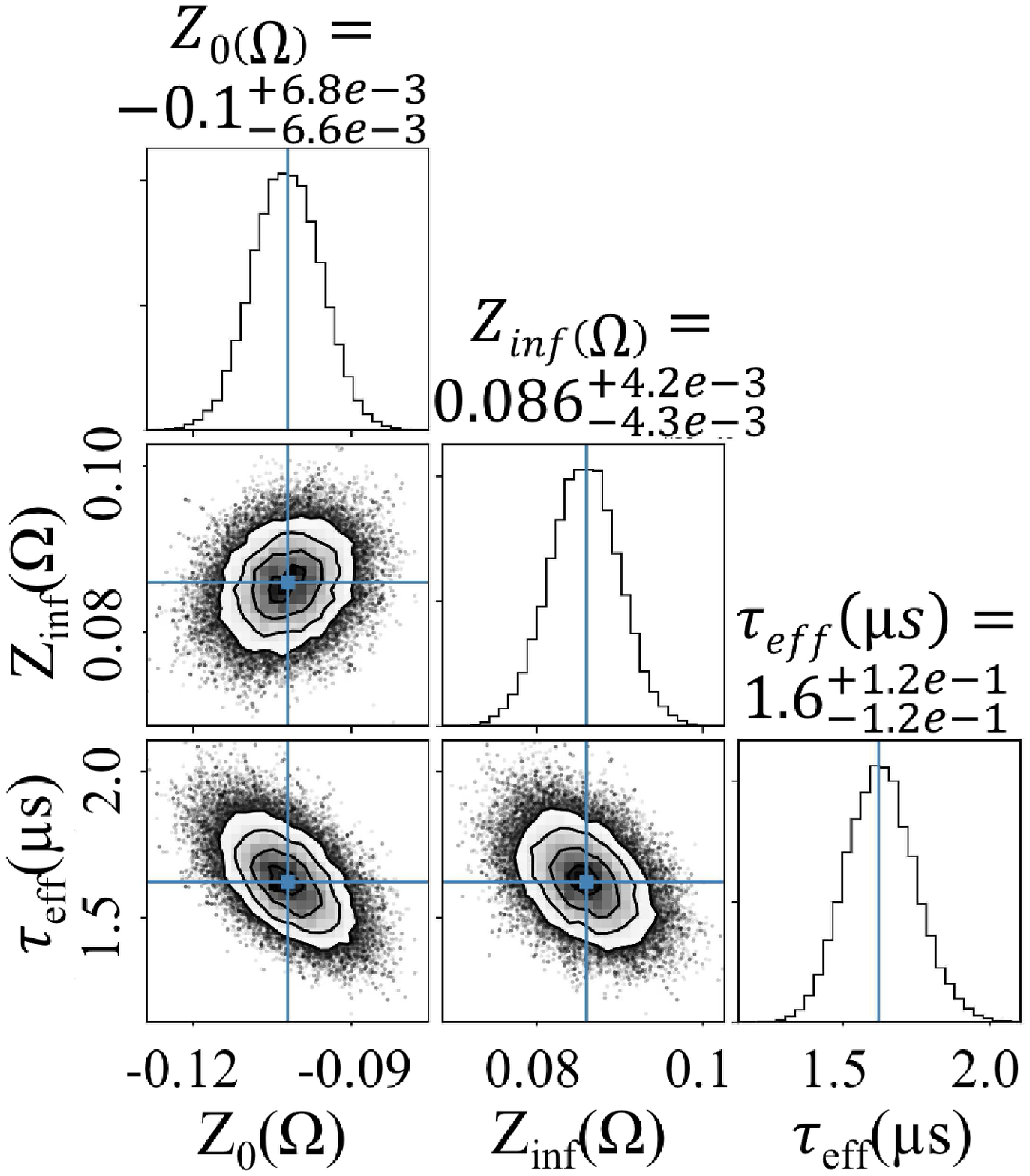}\label{fig:corner}}\\
	
	\subfloat[]{\includegraphics[width=0.7\linewidth, keepaspectratio]{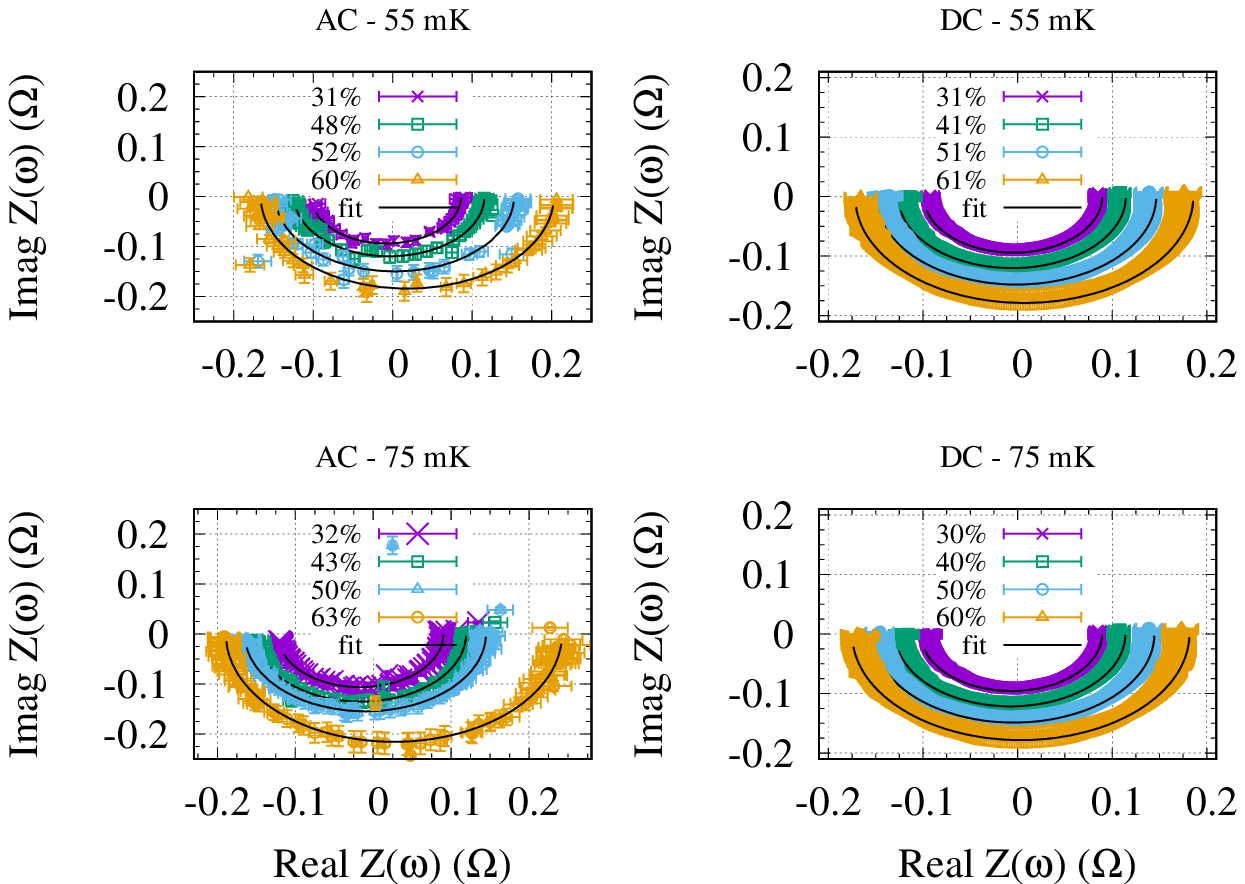} \label{fig:Zmcmc}}\\
\caption{The {\it Top} plot shows the statistical relationship among the fit parameters. On the {\it Bottom} complex impedance measurements at different percentages of R$_N$ ({\it Open symbols}) with related fits ({\it Black lines)} with a bath temperature of 55 mK ({\it top graphs}) and 75 mK ({\it Bottom graphs}) in AC and DC bias respectively. (Color figure online.)}
\label{fig:MCMC}
\end{center}
\end{figure}

We use IV curves, complex impedance and noise measurements in both the setups to characterise and compare the behaviour of our TES. In general the complex impedance is the measurement that needs some additional explanation. All the details about the complex impedance measurements and related calibration can be found in Ref. \cite{taralli} for AC and in Ref. \cite{carlos} for DC case. In this section we give some detail on the Markov Chain Monte Carlo (MCMC) fitting method used to fit the complex impedance measurements.\\
A common reason to use the MCMC method is that it would be useful to marginalise over some parameters and find an estimate of the posterior probability function for others.  For this purpose we first define the fit function $Z_{TES}=Z_{inf}+(Z_{inf}-Z_0)\frac{1}{-1+i\omega\tau_{eff}}$ which is derived from the one-body model that seems to be enough to describe our TES, based on previous measurements of such devices \cite{taralli}. This is our likehood function, where Z$_0$ and Z$_{inf}$ are the low-frequency and the high-frequency limit of the impedance, respectively and $\tau_{eff}$ is the effective time constant of the detector. Afterwards we define their prior probability supposing they are described by normal distribution and their posterior probability that is a conjunction between the prior probability and the likehood function.  We began sampling our parameter space using walkers (much more than twice of the number of parameters being varied during the fit)  starting from a tiny Gaussian ball around the maximum likelihood result obtained from a standard fit with the common method of least squared \cite{mackay}.
The plot in Fig. ~\ref{fig:corner} shows all the one and two dimensional projections of the posterior probability distributions of our parameters. This can quickly demonstrate all of the covariances between parameters and shows the standard deviation for each of them. The analysis showed in Fig. ~\ref{fig:corner} refer to impedance measurement in AC bias at 31\% of the transition with T$_{bath}$=55 mK. In Fig. ~\ref{fig:Zmcmc} there are some of the complex impedance measurement at different bias points in AC and DC case and related fit both at 55 mK (Top) and 75 mK (Bottom). This procedure has been applied to all the bias points for both the setups.

\section{Comparison}

\begin{figure}[htbp]
\begin{center}
\includegraphics[width=0.6\linewidth, keepaspectratio]{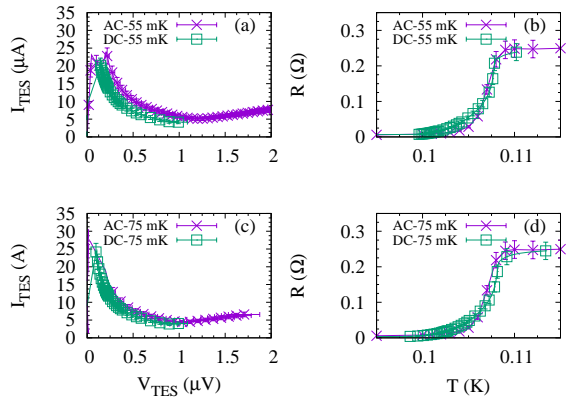}
\caption{IV and extrapolated RT curves of the pixel in AC ({\it Purple crosses}) and DC case ({\it green squared}) at 55 ({\it top graphs}) and 75 mK ({\it Bottom graphs}). (Color figure online).}
\label{fig:IV}
\end{center}
\end{figure}

\begin{figure}[htbp]
\begin{center}
\includegraphics[width=0.6\linewidth, keepaspectratio]{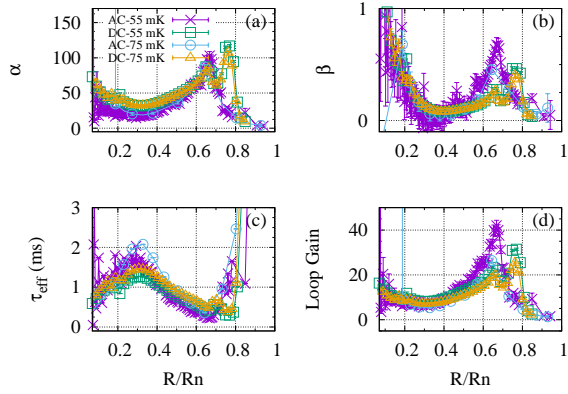}
\caption{Parameters derived from impedance measurements: $\alpha$ ({\it a}), $\beta$ ({\it b}), $\tau_{eff}$ ({\it c}), and loop gain $\mathcal{L}$ ({\it d}) ({\it Open symbols}) as a function of the bias point expressed as R/R$_N$. Lines serve no other purpose to guide the eye. (Color figure online.)}
\label{fig:Zpar}
\end{center}
\end{figure}

\begin{figure}[htbp]
\begin{center}
\includegraphics[width=0.6\linewidth, keepaspectratio]{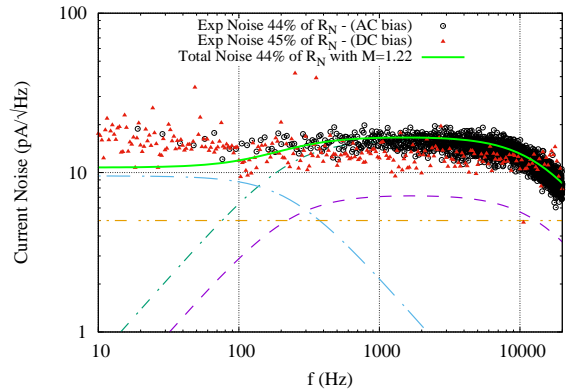}
\caption{Noise measurements of TES20 at 44\% and 45\% of R$_N$ for AC and DC bias, respectively. Noise contributions: SQUID noise ({\it yellow dot-dot-dash line}), Johnson noise ({\it  purple dashed line}), Excess Johnson noise ({\it green dot dash line}) and phonon noise ({\it blue dot-long dash line}). The cutoff at frequencies above 10 kHz in AC bias is due to the use of a band-pass filter to avoid interference with the neighbouring pixel. (Color figure online.)}
\label{fig:noise}
\end{center}
\end{figure}

IV curves have been measured at T$_{bath}$  of 55 mK and 75 mK (Fig. ~\ref{fig:IV}a and \ref{fig:IV}c, respectively) and the corresponding extrapolated RT curves are shown in Fig. ~\ref{fig:IV}b and \ref{fig:IV}d. IV curves point out a small offset between the two curves, probably due to a residual magnetic field that was not possible to be canceled in the DC setup.\\
From the fitting parameters of the complex impedance measurements we can derive the sensitivity of the TES resistance on the temperature $\alpha$ and on the current $\beta$ and the loop gain at low frequency $\mathcal{L}$. We use the following equations: $\beta=\frac{Z_{inf}}{R}-1$, $\mathcal{L}=\frac{C}{\tau_{eff}G}-1$ and $\alpha=\frac{\mathcal{L}GT}{P}$ where R is the TES resistance at the specific bias point, G is derived from the P(T) curve, T is the TES temperature, P is the Joule heat dissipated in the TES at that bias point and the total heat capacity C is the sum of the C$_{ABS}$ (1.18 pJ/K) and the C$_{TES}$ (0.02 pJ/K) \cite{taralli}.\\
In Fig. ~\ref{fig:Zpar} are shown the derived parameters. The values of the key parameters $\alpha$ and $\beta$ in the AC are smaller and the shape seems to be smoothed out compared to the DC. Looking at the curve of  $\alpha$ in the AC, we can see of course the large peak around 68\% of the transition but only an hint of the second one around 58\%. This small peak is a bit more evident in the shape of $\beta$. In the DC those are clearly much more evident and their values are larger as already said. It is clear that there is a shift between AC and DC and this confirm our previous guess about  the presence of a small magnetic field in the DC setup. It has been already demonstrated \cite{steve} that any peak of  $\alpha$ and $\beta$ can appear at different R/R$_N$ for different applied magnetic fields in TESs with normal metal structures. Moreover, we can remark that the behaviour of $\alpha$ and $\beta$ between AC and DC bias start diverging at lower bias point in the transition.\\
One can use the parameters obtained from the complex impedance to model the detector noise. In Fig. ~\ref{fig:noise} the measured noise spectra are shown at 44\% and 45\% of R$_N$ for AC and DC bias, respectively and the results from the model, using the AC bias parameters, are over-plotted. The model noise contributions are: phonon noise, TES Johnson noise, excess Johnson noise and SQUID noise. Those noise sources describe very well the noise observed at frequencies higher than 100 Hz and it looks slightly lower in the DC case.  In the frequency range where the Johnson noise is dominant there is an excess noise, which is quantified as M times the Johnson noise and introduced by this factor M \cite{ullom}.
Considering the predicted energy resolution $2.355\sqrt{4k_bT_0^2\times\frac{C}{\alpha}\sqrt{nF(1+2\beta)(1+M^2)}}$ for the same bias point of the noise spectra shown in Fig. ~\ref{fig:noise},  we obtain 3.3 eV in AC against the 2.8 eV in DC. This discrepancy of $\sim$10\%  could be larger when the behaviour of the detector under AC begins to differ from the DC. This is in line with the assumption that detectors with high saturation power and high normal resistance, or biased at higher bias points in the transition, show generally small or negligible Josephson current under ac bias,  mitigating the weak-link behaviour in ac-biased detectors.\\

\vspace{-1cm}

\section{Conclusion}
\vspace{-0.1cm}

We have performed a comparison of the performance of a single pixel TES microcalorimeter under DC and AC bias (f$_c$ = 3.5 MHz), by means of IV curves, complex impedance and noise measurements. The behaviour of the detector under AC begins to differ from the DC at working points lower in the transition. In the AC, values of $\alpha$ and $\beta$ are lower and their shape appear to be smoothed out, especially in presence of peaks. A better analysis, including the error estimation on the parameters obtained from complex impedance data has been presented to guarantee a fair comparison.\\
SRON is currently developing high aspect ratio TiAu TES, with a thicker bilayer and without normal metal structures, high normal  resistance and high saturation power to accomplish the goal of having high performance detectors under AC bias\cite{taralli2}. In the future, we are planning for these new pixels extensively AC and DC bias experiments including X-ray energy resolution measurement at different bias points to better understand the interaction between these devices and the voltage bias readout system.\\

\begin{acknowledgements}
This work is partly funded by European Space Agency (ESA) and coordinated with other European efforts under ESA CTP contract ITT AO/1-7947/14/NL/BW. It has also received funding from the European Union's Horizon 2020 Programme under the AHEAD (Activities for the High-Energy Astrophysics Domain) project with grant agreement number 654215. CSIC work financed by the Spanish Ministerio de Ciencia, Innovaci\'on y Universidades-MICINN (projects ESP2016-76683-C3-2-R and RTI2018-096686-B-C22).  Personnel from ICMAB acknowledge financial support from MINECO, through the 'Severo Ochoa
\end{acknowledgements}

\begin{acknowledgements}

  '  Programme for Centres of Excellence in R\&D (SEV- 2015-0496). 

\end{acknowledgements}



\begin{thebibliography}{99}
\small
\bibitem{xifu}
F. Pajot, D. Barret, T. Lam-Trong, J.-W. den Herder et al., {\it J. Low Temp. Phys.} \textbf{193}, 901 (2018). DOI:  https://doi.org/10.1007/s10909-018-1904-5

\bibitem{athena}
European Space Agency (ESA),''ATHENA Mission Summary,''Available: http://sci.esa.int/athena/59896-mission-summary.

\bibitem{doriese}
W. B. Doriese et al., {\it J. Low Temp. Phys.} \textbf{184}, 389 (2016). DOI: https://doi.org/10.1007/s10909-015-1373-z

\bibitem{gottardi}
L. Gottardi, J. van der Kuur, P. A. J. de Korte, R. Den Hartog, B. Dirks, M. Popescu, H. F. C. Hoevers, M. Bruijn, M. Parra Borderias, and Y. Takei, {\it AIP Conference Proceedings} \textbf{1185}. DOI: 10.1063/1.3292399.

\bibitem{pourya}
P. Khosropanah, E. Taralli, L. Gottardi, K. Nagayoshi, M. Ridder, M. Brujin, and J.-R. Gao, {\it Proceedings volume 9144 SPIE Atronomical Telescopes + Insrumentation,
Space Telescopes and Instrumentation 2018: Ultraviolet to Gamma Ray}, Austin, Texas, USA, 10-15 June 2018, edited by J.-W. A. den Herder, K. Nakazawa, and S. Nikzad.

\bibitem{Qlc}
L. Gottardi, J. van der Kuur, M. Bruijn, A. van der Linden, M. Kiviranta, H. Akamatsu, R. den Hartog, and K. Ravensberg, {\it J. Low Temp. Phys.} {\textbf 194}, 370, (2018). DOI: https://doi.org/10.1007/s10909-018-2085-y

\bibitem{taralli}
E. Taralli, P. Khosropanah, L. Gottardi, K. Nagayoshi, M. L. Ridder, M. P. Bruijn, and J. R. Gao, {\it AIP Advances} {\textbf 9}, 045324, (2019), DOI:10.1063/1.5089739

\bibitem{carlos}
C. Pobes, L. F\`abrega, A. Cam\'on, N. Casa\~n-Pastor, P. Strichovanec, J. Ses\'e, J. Moral-Vico, R. M. J\'audenes Calleja, {\it IEEE Trans. Appl. Supercond.} \textbf{27}, 2101505, (2017), DOI: 10.1109/TASC.2016.2637337

\bibitem{mackay} D. J. C. MacKay, {\it Information Theory, Inference, and Learning Algorithms.} Cambridge, U.K.: Cambridge Univ. Press, 2003.
\bibitem{steve}
S.J. Smith, J. S. Adams, C. N. Bailey, S. R. Bandler, S. E. Busch, J. A.Chervenak, M. E. Eckart, F. M. Finkbeiner, C. A. Kilbourne, R. L. Kelley, S.-J. Lee, J.-
P. Porst, F. S. Porter, and J. E. Sadleir, {\it J. Appl. Phys.} \textbf{114} 074513 (2013). . DOI: https://doi.org/10.1063/1.4818917

\bibitem{ullom}
N. Ullom, J. A. Beall, W. B. Doriese, W. Duncan, S. L. Ferreira, G. C. Hilton, K. D. Irwin, C. D. Reintsema, and L. R. Vale, {\it Appl. Phys. Lett.} \textbf{87}, 194103 (2005). DOI: https://doi.org/10.1063/1.2061865

\bibitem{taralli2}
E. Taralli, L. Gottardi, K. Nagayoshi, M. Ridder, S. Visser, P. Khosropanah, M. Bruijn and J.R. Gao, {\it J. Low Temp. Phys.}. This Special Issue (2019).

\end{thebibliography}
\end{document}